\documentclass[runningheads]{llncs}

\usepackage{graphicx}
\usepackage{todonotes}
\usepackage{subcaption}
\usepackage{hyperref}
\usepackage{xurl}

\begin{document}
\begin{sloppypar}

\title{Archiverse: an Approach for \\Immersive Cultural Heritage}

\titlerunning{Archiverse}

\author{
Wiesław Kopeć \inst{1,4}\orcidID{0000-0001-9132-4171} \and
Anna Jaskulska\inst{4}\orcidID{0000-0002-2539-3934} \and
Władysław Fuchs\inst{2,3}\orcidID{0000-0002-9069-6181}\and
Wiktor Stawski \inst{1}\orcidID{0000-0001-8950-195X} \and
Stanisław Knapiński \inst{1}\orcidID{0009-0006-0524-2545} \and
Barbara Karpowicz\inst{1}\orcidID{0000-0002-7478-7374} \and
Rafał Masłyk\inst{1}\orcidID{0000-0003-1180-2159}
}

\authorrunning{Kopeć et al.}

\institute{XR Center, Polish-Japanese Academy of Information Technology
 \url{https://xrc.pja.edu.pl}
\and
University of Detroit Mercy
\and
Volterra-Detroit Foundation
\and
Kobo Association
}

\maketitle              

\begin{abstract}

Digital technologies and tools have transformed the way we can study cultural heritage and the way we can recreate it digitally. Techniques such as laser scanning, photogrammetry, and a variety of Mixed Reality solutions have enabled researchers to examine cultural objects and artifacts more precisely and from new perspectives. In this part of the panel, we explore how Virtual Reality (VR) and eXtended Reality (XR) can serve as tools to recreate and visualize the remains of historical cultural heritage and experience it in simulations of its original complexity, which means immersive and interactive. Visualization of material culture exemplified by archaeological sites and architecture can be particularly useful when only ruins or archaeological remains survive. However, these advancements also bring significant challenges, especially in the area of transdisciplinary cooperation between specialists from many, often distant, fields, and the dissemination of virtual immersive environments among both professionals and the general public.


\end{abstract}

\section{Rationale}

Digital recreation or reconstruction and immersive visualization of lost cultural heritage architectural objects poses significant challenges that require the collaboration of specialists from a wide range of disciplines. This part of the discussion panel is based on recent research advancements in creating high-quality VR environments using data from architectural BIM (Building Information Modeling) systems as a way of presenting digital reconstructions of cultural heritage conducted as part of the Archiverse framework project. This project stems from the previous experience in virtual and immersive environments by the HASE (Human Aspects in Science and Engineering) research group~\cite{karpo2022avatars,kopec2023human,pochwatko2023invisible,pochwatko2023well,schudy2023} by the Living Lab Kobo and benefits from research and activities of Wladyslaw Fuchs~\cite{fuchs2019geometric1,fuchs2019geometric2,fuchs2021study} and Volterra-Detroit Foundation, in particular based on the roman theater in Volterra use-case.

The Roman theater in Volterra was discovered relatively late, in the mid-20th century, thanks to the efforts of Enrico Fiumi, who initiated archaeological excavations. Excavations under his direction were conducted between 1950 and 1953.~\cite{fiumi1955volterra} Subsequently, the work was taken over by the state archaeological services. At the turn of the 1960s and 1970s, anastylosis was carried out, i.e., the reassembly of columns and other architectural elements from the ruined building, supplemented by the partial reconstruction of a fragment of one of the walls.
The theater in Volterra has survived in relatively good condition. Compared to other similar structures, the stage building of the theater is exceptionally well-preserved.~\cite{inghirami1977teatro}

\begin{figure}
    \centering
    \includegraphics[width=0.8\linewidth]{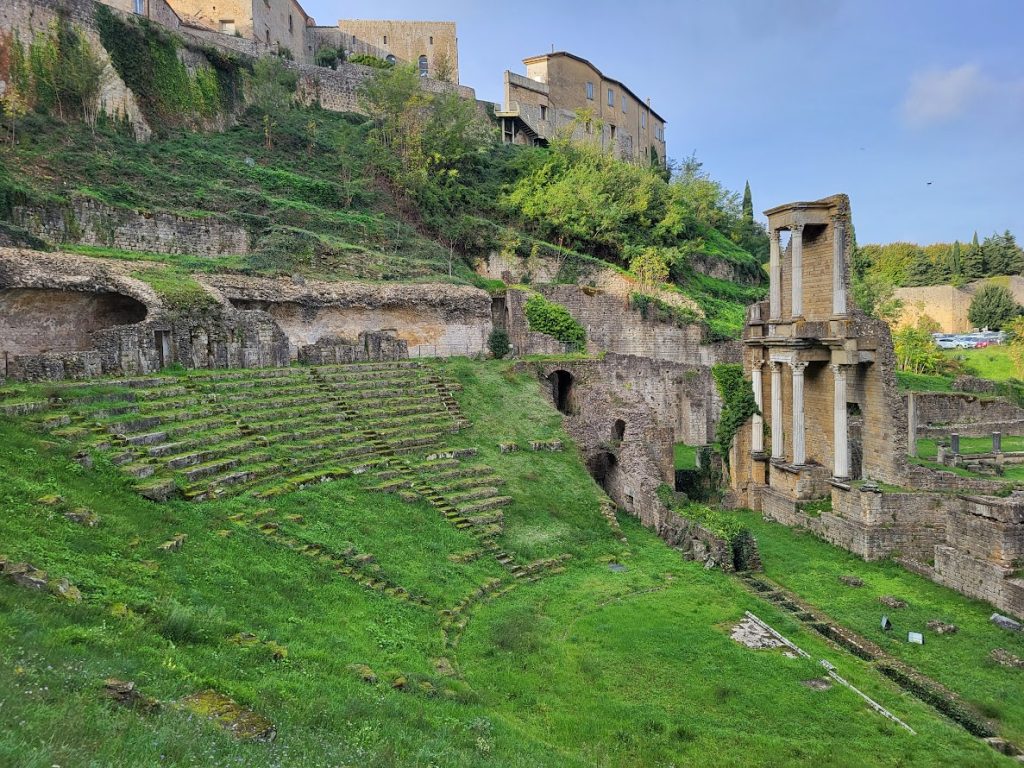}
    \caption{Roman theater in Volterra (source: own elaboration)}
    \label{fig:theater}
\end{figure}

In October 2016, during a workshop organized by the Volterra-Detroit Foundation with the participation of Case Technologies and Autodesk, data acquisition of the remains of the Roman theater was carried out. Using modern methods involving laser scanning, photogrammetry, and drone imagery, very precise data was obtained. Detailed analyzes were conducted by Władysław Fuchs, a professor at the University of Detroit Mercy (School of Architecture and Community Development) and head of the Volterra-Detroit Foundation, who has been researching methods of designing and constructing buildings in ancient Rome for years. Fuchs' research revealed certain features of the theater in Volterra (see Figure \ref{fig:theater}) that had previously been overlooked or underestimated.~\cite{fuchs2021study}

To continue research on spatial composition, Fuchs also prepared a three-dimensional model of the theater. The goal of the model was not to create a digital replica of the original structure - some architectural details were simplified.
Fuchs discovered that the building is characterized by a very regular and clear composition. Its architect skillfully used complex and unusual geometric arrangements. The auditorium is based on a heptagon geometry and its multiplication in the form of a regular polygon with an odd number of 21 sides. When designing the stage space, the architect departed from the concept of a continuous facade. Three independent buildings, separated by narrow streets, create the illusion of an urban landscape.~\cite{fuchs2021study}

\begin{figure}
    \centering
    \includegraphics[width=0.8\linewidth]{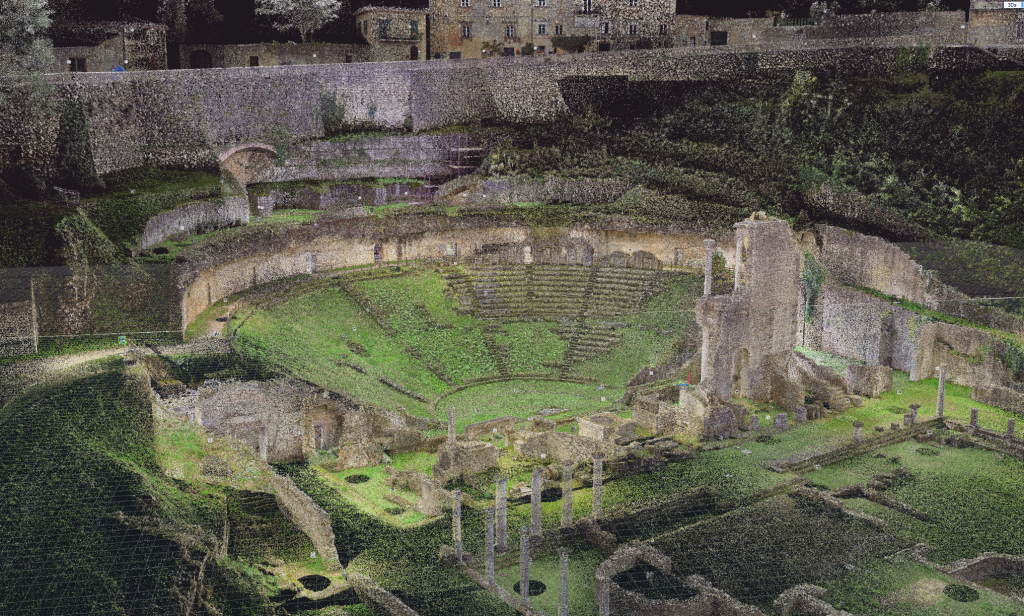}
    \caption{Roman theater in Volterra - point cloud based on laser scan (source: own elaboration)}
    \label{fig:pointCloud}
\end{figure}

Fuchs also prepared a 3D digital reconstruction showing what the theater might have looked like at the turn of the 2nd and 3rd centuries AD. Despite the significant amount of research on which the theater's space is based, the project has varying levels of confidence regarding the individual elements. 

In order to conduct further experiments in an environment that allows interaction with the space of ancient architecture in a way that is similar to reality and to test the principles of the framework within the project's use case, a virtual, immersive reconstruction of the theater in Volterra was developed based on Fuchs' BIM models.

\section{Focus}
During the panel, we explore how rapidly developing technologies, including Virtual Reality (VR) and eXtended Reality (XR), can serve as tools to recreate and visualize the remains of historical cultural heritage for a better experience of lost architectural heritage in the multidisciplinary research context. Researchers with expertise and practice in a wide range of specializations, from architecture and design, to immersive real time environment development, shared their experiences with a roman theater in Voltera use-case.
Moreover, the participants of the conference also had an opportunity to visit the exhibition taking place in the PJAIT gallery - an interactive presentation of an immersive reconstruction of the roman theater in Volterra in VR headsets. The virtual environment was accompanied by a special multimedia presentation of the cultural context of the project and the main principles of the Archiverse Framework.

\subsection{Discussion Points}

\paragraph{Aspect 1: How are new technologies advancing research methods in cultural heritage studies? What are the advantages of VR technologies in visualizing lost architectural heritage?}

\paragraph{Aspect 2: What challenges are faced when using VR to visualize lost architectural heritage?}

\paragraph{Aspect 3: What are the challenges faced by teams composed of members from various disciplines, backgrounds, and expertise?}

\section{Organizers}

\subsection{Panelists}

\paragraph{Władysław Fuchs}
Received his Master of Architecture degree in 1987 from the Warsaw Institute of Technology. In 1994, he completed and defended his Ph.D. dissertation at the same institution. He has been awarded the Warsaw Institute of Technology President's Award. From 1987 until 1991 he worked full-time in the Free-hand Drawing Department of the School of Architecture in Warsaw. Since 1991 he has been working continuously at the School of Architecture, University of Detroit Mercy. 

\paragraph{Wiktor Stawski}
Head of the XR\_Students' Club and a researcher at the XR\_Center's laboratory. Passionate about 3D printing, Virtual Reality (VR), and eXtended Reality (XR). With a background in Interior Architecture, he has a keen interest in architecture and exhibitions. He presented a proposal for the reconstruction of an ancient Roman theater in Volterra using real-time engines at the MIDI 2024 conference. Additionally, he participated in an analog space mission at a simulation facility in collaboration with an ESA engineer. His work bridges the gap between traditional architecture and modern digital tools, showcasing innovative approaches to architectural implementation.

\paragraph{Wiesław Kopeć}
Computer scientist, research and innovation team leader, associate professor at the Computer Science Faculty of Polish-Japanese Academy of Information Technology (PJAIT). Head of the XR Center PJAIT and the XR Department. He is also a seasoned project manager with a long-term collaboration track with many universities and academic centers, including the University of Warsaw, the SWPS University, National Information Processing Institute, and the institutes of Polish Academy of Sciences. He co-founded the transdisciplinary HASE research group (Human Aspects in Science and Engineering) and distributed LivingLab Kobo.

\subsection{Researchers network}
We would like to thank the many people and institutions gathered by the Kobo Living Lab and the HASE Research Group to allow for this collaboration and research. First, we thank all the members of HASE research group (Human Aspects in Science and Engineering) and the Living Lab Kobo for their support. In particular, the members of XR Center Polish-Japanese Academy of Information Technology (PJAIT) as well as its students' club, supporting Volterra preservation efforts. We would also like to thank the members of the Volterra-Detroit Foundation, in particular Giulia Munday, Marco Bruchi, Brey Tucker, Paul F. Aubin, and Mark E. Dietrick.

\section{Discussion and Conclusions}

Archaeological documentation has long been supplemented by various forms of visual materials. Initially, these took the form of drawings depicting plans, sections, and preserved architectural details. The introduction of photography into archaeology seemingly increased the level of realism and provided a direct link between the materials analyzed during excavations and their visualization. However, the trap of the illusion of objectivity in photography can also be noticeable in this case. Many variables depend on the subjective decisions of the documentary photographer, who makes decisions about, among other things, the choice of frame, lighting, and time of day for taking the photo, as well as the equipment and processing of the finished image. In the case of the theater in Volterra, researchers analyzing the anastylosis carried out at the turn of the 1960s and 1970s mention problems with the readability of the preserved photographic documentation, despite the fact that it was conducted in modern times, under the supervision of state archaeological services. The implementation of state-of-the-art technologies such as 3D laser scanning (see Figure \ref{fig:pointCloud}), drone imaging, and photogrammetry in archaeological research is revolutionizing data collection. These methods allow for the acquisition of highly detailed and objective data, enabling a more comprehensive understanding of archaeological sites. They are being applied to both re-examine existing sites, like the Roman theater in Volterra, and to document ongoing excavations, such as the amphitheater in Volterra, which was unearthed in 2015 and has been studied intermittently since 2019.

Despite the above-mentioned technological advancements, digital representations of architecture are mainly reliant on static two-dimensional depictions, such as 3D models presented on flat screen or in printed form. This approach fails to capture the complex dynamic of reality, which influences human perception of architecture and provides an incomplete and potentially misleading understanding of spatial relationships between individuals and their surroundings. Key factors like construction techniques, spatial configuration and scale, lighting and decorative details are inherently complex and cannot be adequately studied or understood in two dimensions. To gain a more accurate understanding of how buildings were constructed, what their purpose was, and how they were used by real people, researchers should prioritize the creation and interactive exploration of 3D simulations that recreate the original complexity of these structures.~\cite{sanders2016more} 

\begin{figure}
    \centering
    \includegraphics[width=0.8\linewidth]{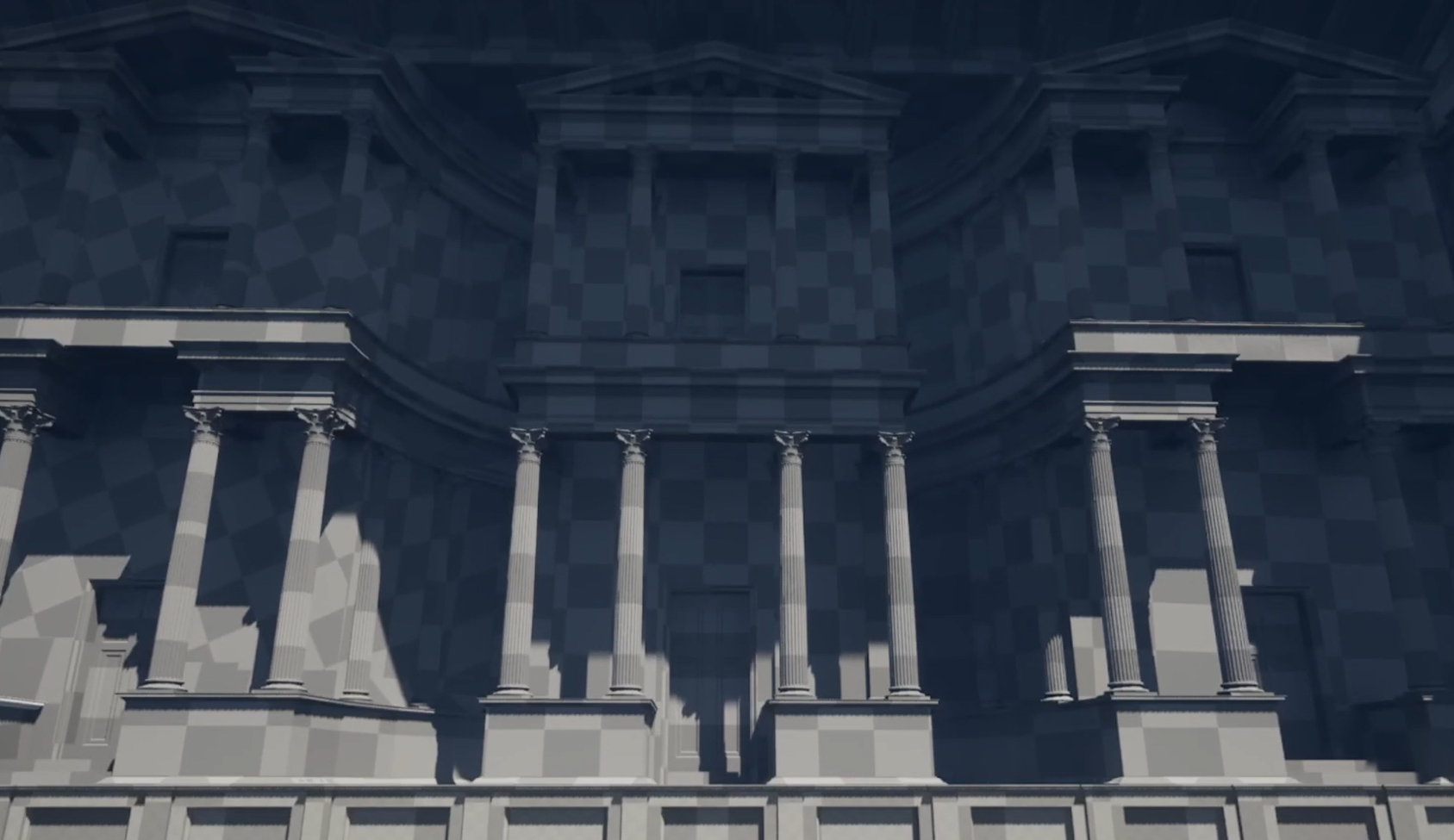}
    \caption{Roman theater in Volterra - graybox prototype (source: own elaboration)}
    \label{fig:graybox}
\end{figure}

\begin{figure}
    \centering
    \includegraphics[width=0.8\linewidth]{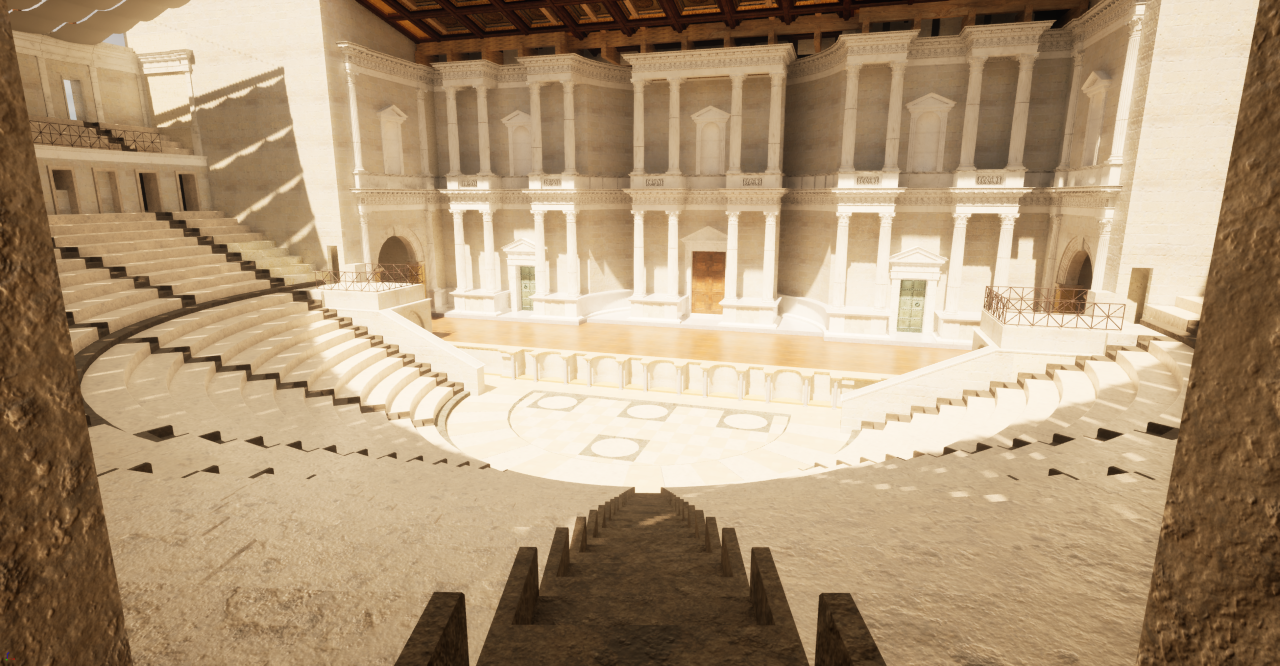}
    \caption{Roman theater in Volterra - immersive visualisation (source: own elaboration)}
    \label{fig:immersiveVisualisation}
\end{figure}

Developing an immersive virtual experience that accurately represents lost architectural heritage requires a five-step process: data acquisition and modeling (see Figure \ref{fig:pointCloud}), conversion for real-time environments (see Figure \ref{fig:graybox}), development of immersive environment (see Figure \ref{fig:immersiveVisualisation}), and composition of narrative interactive experience. The process of capturing source data for these environments involves utilizing 3D scanners, photogrammetry, and drone-based data acquisition. The acquired data should be analyzed and processed in a cross-disciplinary research context based on sources, literature reviews, and research hypotheses. 

Such a comparative, cross-disciplinary research process is necessary for proper interpretation of the data gathered and provides the foundation for reconstruction concepts and actual architectural modeling using various BIM methods and tools. Such models are the basis for further real-time immersive visualizations and need to be converted for real-time immersive purposes. The level of automation of this process varies depending on the chosen real-time graphics engine for the immersive environment. Additionally, depending on the selected engine, AI-powered tools can assist in various aspects of VR environment preparation. Despite the progress in automation, real time graphics constrains and limitations are still one of the main challenge in the process of creation of immersive virtual visualizations. Ultimately, to create a compelling VR presentation, a narrative part must be prepared. The high level of immersion may require not only technical conversion of the reconstructed architectural elements, but also elements of the hypothetical scene, including avatars and interactive elements incorporated to enhance the viewer's immersive experience.

It should be noted that this process necessitates interdisciplinary collaboration between experts from a wide range of fields. This is a complex undertaking that often requires the guidance of experienced facilitators with a cross-disciplinary background. Knowledge transfer must occur not only between technologists and humanities scholars, but also among specialists at different stages of the project, such as data acquisition experts, BIM modelers, real-time environment developers, and researchers from various disciplines, including archaeology, architecture, history, art history, and cultural studies.~\cite{kopec2023co}

\section*{Acknowledgments}
The project \textit{Creating high-quality VR environments using data from architectural BIM systems as a way of presenting digital reconstructions of cultural heritage}, which made Archiverse Framework and Volterra VR Use Case research possible, received funding from the National Recovery Plan. It was funded by the European Union - NextGenerationEU (grant number 1475/KPO. GRANTY/NIMiT/2024).
\bibliographystyle{splncs04}
\bibliography{bibliography}

\end{sloppypar}
\end{document}